\documentclass[twocolumn,showpacs,preprintnumbers]{revtex4}%
\usepackage{amsmath}
\usepackage{graphicx}
\usepackage{dcolumn}
\usepackage{bm}
\usepackage{amsfonts}
\usepackage{amssymb}%
\providecommand{\U}[1]{\protect\rule{.1in}{.1in}}
\begin{document}
\title{Coherent population oscillation from a Fermi atom-molecule dark state}
\author{$^{1}$Andrew Robertson, Lei Jiang$^{2}$, Han Pu$^{2}$, Weiping Zhang$^{3}$,
and Hong Y. Ling$^{1,\dag}$}
\affiliation{$^{1}$Department of Physics and Astronomy, Rowan University, Glassboro, New
Jersey, 08028-1700, USA}
\affiliation{$^{2}$Department of Physics and Astronomy, and Rice Quantum Institute, Rice
University, Houston, TX 77251-1892, USA}
\affiliation{$^{3}$Key Laboratory of Optical and Magnetic Resonance Spectroscopy (Ministry
of Education), Department of Physics, East China Normal University, Shanghai
200062, P. R. China}

\begin{abstract}
We show that a robust macroscopic atom-molecule dark state can exist in
fermionic systems, which represents a coherent superposition between the
ground molecular BEC and the atomic BCS paired state. We take advantage of the
tunability offered by external laser fields, and explore this superposition
for demonstrating coherent oscillations between ground molecules and atom
pairs. We interpret the oscillation frequencies in terms of the collective
excitations of the dark state.

\end{abstract}
\date{\today }

\maketitle

Association of ultracold atom pairs into diatomic molecules via Feshbach
resonance \cite{tiesinga93} or photoassociation \cite{thorsheim87}, has made
it possible to create coherent superpositions between atomic and molecular
species at macroscopic level. This ability is the key to applications that
employ the principle of the double pulse Ramsey interferometer \cite{ramsey85}
for observing coherent population oscillations between atoms and molecules
\cite{donley02,kokkelmans02,Mackie02}. A particular kind of state, the
atom-molecule dark state, has been theoretically proposed \cite{mackie,hong}
and experimentally observed \cite{exp}, where population is trapped in a
superposition between atom pairs and deeply bound molecules in the electronic
ground state. Destructive interference leads to the vanishing population in
the excited molecular level. Such a state is the generalization of the usual
atomic dark state that lies at the heart of many exciting applications,
including electromagnetically induced transparency, slow light propagation and
precision spectroscopy \cite{scully}. So far, the macroscopic atom-molecule
dark state has only been studied in bosonic systems. The purpose of this paper
is to show that, under proper conditions, an atom-molecule dark state also
exists in fermionic systems, but with quite distinct properties compared with
its bosonic counterpart.

To be specific, we consider a homogeneous atom-molecule system where an
excited molecular level $|m\rangle$ is coupled both to a ground molecular
level $|g\rangle$ (bound-bound coupling) by a coherent laser field, and to two
free atomic states of equal population labeled as $\left\vert \uparrow
\right\rangle $ and $\left\vert \downarrow\right\rangle $ (bound-free
coupling) via, for example, a photoassociation laser field. At zero
temperature, bosonic molecules all condense to the zero-momentum state,
whereas fermionic atoms are of multi-momentum modes in nature due to the Pauli
principle, and are thus described by momentum continua of different internal
states. This difference has two important ramifications.

The first one is related to the formation of the dark state. \ As is known,
two necessary ingredients for creating a macroscopic atom-molecule dark state
are the coherence between its components and the generalized two-photon
resonance which, unlike in the linear atomic model, becomes explicitly
dependent on the atomic momentum. \ For bosons at zero temperature, since they
all occupy the same zero-momentum mode, properly tuning the laser frequencies
can make all the bosons satisfy the two-photon resonance simultaneously.
\ However, for fermions, because of the existence of the fermi momentum sea,
the same technique can only render a limited number of atoms with the
\textquotedblleft right\textquotedblright\ momentum to satisfy the two-photon
resonance. \ Hence a macroscopic dark state involving all the particles in the
system does not seem to be possible for fermions. \ This difficulty can be
circumvented when the attractive interaction between atoms of opposite spins
results in a fermionic superfluid state that can be regarded as a condensate
of atomic cooper pairs. As we shall show below, such a fermionic superfluid,
together with the ground molecule condensate, can now form a macroscopic dark
state under the two-photon resonance condition.

The second ramification of the momentum continuum is related to the collective
excitation of the dark state. The excitation spectrum of the fermionic system
is far more difficult to analyze than its bosonic counterpart. The
zero-temperature spectrum of the bosonic system is discrete \cite{hong}. In
contrast, the spectrum of the fermionic system is made up of both a discrete
and a continuous part, and hence can be regarded as the nonlinear analog of
the Fano-Anderson type of models in linear atomic and condensed matter systems
\cite{fano61}. As we demonstrate later, this analogy significantly simplifies
our understanding of the excitation spectrum while at the same time enables us
to gain profound insights into the dynamical properties of the fermionic dark state.

Let us begin with the mean-field Hamiltonian \cite{gennes89} written in the
frame rotating at the laser frequency:
\begin{align}
\hat{H}  &  =\sum_{\mathbf{k},\sigma}\epsilon_{k}\hat{a}_{\mathbf{k},\sigma
}^{\dag}\hat{a}_{\mathbf{k},\sigma}+\nu_{0}\hat{b}_{m}^{\dag}\hat{b}%
_{m}+\left(  \delta_{0}+\nu_{0}\right)  \hat{b}_{g}^{\dag}\hat{b}%
_{g}\nonumber\\
&  -\sum_{\mathbf{k}}\varphi_{k}\left(  \Delta\,\hat{a}_{\mathbf{k},\uparrow
}^{\dag}\hat{a}_{-\mathbf{k},\downarrow}^{\dag}+h.c\right)  +\frac{\Omega_{0}%
}{2}\left(  \hat{b}_{m}^{\dag}\hat{b}_{g}+h.c\right) \nonumber\\
&  +\frac{1}{\sqrt{V}}\sum_{\mathbf{k}}g\varphi_{k}\left(  \hat{b}_{m}\hat
{a}_{+\mathbf{k},\uparrow}^{\dag}\hat{a}_{-\mathbf{k},\downarrow}^{\dag
}+h.c\right)  , \label{H}%
\end{align}
where $\hat{a}_{\mathbf{k},\sigma}$ is the annihilation operator for an atom
of spin $\sigma(=\uparrow$ or $\downarrow)$, having momentum $\hbar\mathbf{k}$
and kinetic energy $\epsilon_{k}=\hbar^{2}k^{2}/2m$, $\hat{b}_{m,g}$ the
annihilation operator for a bosonic molecule in state $\left\vert
m\right\rangle $ or $\left\vert g\right\rangle $. We have neglected the
Hartree mean-field potential as it is usually weak for typical parameters.
Here, $V$ is the system volume, $\delta_{0}$ and $\Omega_{0}$ ($\nu_{0}$ and
$g$) are respectively the detuning and coupling strength of the bound-bound
(bound-free) transition, $\varphi_{k}=\exp\left[  -k^{2}/(2K_{c}^{2})\right]
$ is the regularization function providing momentum cutoff, and $\Delta
=-U\sum_{\mathbf{k}}\varphi_{k}\left\langle \hat{a}_{-\mathbf{k},\downarrow
}\hat{a}_{\mathbf{k},\uparrow}\right\rangle /V$ is the gap parameter. The
collisional interaction potential between atoms of opposite spins and the
atom-molecule coupling are given by $U\left(  \mathbf{k}-\mathbf{k}^{\prime
}\right)  $ $=U\varphi_{k}\varphi_{k^{\prime}}$ and $g\left(  \mathbf{k}%
\right)  =g\varphi_{k}$, respectively, where $U$ and $g$ are momentum
independent. Evidently, Eq. (\ref{H}) preserves the total atom number
$N=2(\langle\hat{b}_{m}^{\dag}\hat{b}_{m}\rangle+\langle\hat{b}_{g}^{\dag}%
\hat{b}_{g}\rangle)+2\sum_{\mathbf{k}}\langle\hat{a}_{\mathbf{k},\uparrow
}^{\dag}\hat{a}_{\mathbf{k},\uparrow}\rangle$.

The dynamics of the system is governed by the Heisenberg equations of motion
for operators. By replacing bose operator $\hat{b}_{m,g}$ with the related
c-number $c_{m,g}=\langle\hat{b}_{m,g}\rangle/\sqrt{V}$ and fermi operator
$\hat{a}_{\mathbf{k},\sigma}\left(  t\right)  $ with $u_{k}\left(  t\right)  $
and $v_{k}\left(  t\right)  $ through the Bogoliubov transformation
\begin{equation}
\left[
\begin{array}
[c]{c}%
\hat{a}_{\mathbf{k},\uparrow}\left(  t\right) \\
\hat{a}_{-\mathbf{k},\downarrow}^{\dag}\left(  t\right)
\end{array}
\right]  =\left[
\begin{array}
[c]{cc}%
u_{k}^{\ast}\left(  t\right)  & v_{k}\left(  t\right) \\
-v_{k}^{\ast}\left(  t\right)  & u_{k}\left(  t\right)
\end{array}
\right]  \left[
\begin{array}
[c]{c}%
\hat{\alpha}_{\mathbf{k},\uparrow}\\
\hat{\alpha}_{-\mathbf{k},\downarrow}^{\dag}%
\end{array}
\right]  \label{bogoliubov}%
\end{equation}
with $\left\vert u_{k}\left(  t\right)  \right\vert ^{2}+\left\vert
v_{k}\left(  t\right)  \right\vert ^{2}=1$ and $\hat{\alpha}_{\mathbf{k}%
,\sigma}$ being fermi quasiparticle operators, we obtain the following
equations
\begin{subequations}
\label{mean-field equations}%
\begin{align}
i\hbar\frac{dc_{m}}{dt}  &  =\nu_{0}c_{m}+\frac{\Omega_{0}}{2}c_{g}-\frac
{g}{U}\Delta,\\
i\hbar\frac{dc_{g}}{dt}  &  =\left(  \delta_{0}+\nu_{0}\right)  c_{g}%
+\frac{\Omega_{0}^{\ast}}{2}c_{m},\\
i\hbar\frac{du_{k}}{dt}  &  =-\epsilon_{k}u_{k}+\varphi_{k}\left(  g^{\ast
}c_{m}^{\ast}-\Delta^{\ast}\right)  v_{k},\\
i\hbar\frac{dv_{k}}{dt}  &  =\epsilon_{k}v_{k}+\varphi_{k}\left(
gc_{m}-\Delta\right)  u_{k},\\
\Delta\left(  t\right)   &  =-\frac{U}{V}\sum_{\mathbf{k}}\varphi_{k}%
u_{k}^{\ast}\left(  t\right)  v_{k}\left(  t\right)  ,
\end{align}
where we have assumed that the state of the system is the quasiparticle vacuum
annihilated by $\hat{\alpha}_{\mathbf{k},\sigma}$.%
%TCIMACRO{\FRAME{ftbpFU}{3.5665in}{5.2053in}{0pt}{\Qcb{(Color online) (a)
%$\left\vert c_{g}^{s}\right\vert ^{2}/n$, $\mu$, and $\Delta^{s}$ as functions
%of $\Omega_{0}$. The dark state solution at $\Omega_{0}=4.1\ E_{F}$ is
%indicated by the vertical line where $\left\vert c_{g}^{s}\right\vert
%^{2}/n=0.067$, $\mu=0.87\ E_{F}$, and $\Delta^{s}=0.17\ E_{F}$. (b)-(d) The
%ground population dynamics where (b) $\nu_{0}=-4.32$ $E_{F}$, (c) $\nu
%_{0}=-2.88\ E_{F}$, and (d) $\nu_{0}=0.00\ E_{F}$. \ The insets are the
%Fourier spectra of the corresponding population dynamics after $t=t_{s}%
%=126.65\hbar/E_{F}.$ We have used the following parameters: $U_{0}%
%=-28.39\,E_{F}\,/k_{F}^{3}$, $g_{0}=-15.68\,E_{F}\,/k_{F}^{3/2}$,
%$n=0.034\,k_{F}^{3}$, and $K_{c}=14.4k_{F}$, where $E_{F}$ and $k_{F}$ are the
%fermi energy and momentum, respectively. }}{\Qlb{Fig:populationsDynamics}%
%}{finaldynamicsnew.eps}{\special{ language "Scientific Word";
%type "GRAPHIC";  maintain-aspect-ratio TRUE;  display "USEDEF";
%valid_file "F";  width 3.5665in;  height 5.2053in;  depth 0pt;
%original-width 2.2459in;  original-height 3.2906in;  cropleft "0";
%filename 'finaldynamicsNew.eps';file-properties "XNPEU";}}}%
%BeginExpansion
\begin{figure}
[ptb]
\begin{center}
\includegraphics[
height=5.2053in,
width=3.5665in
]%
{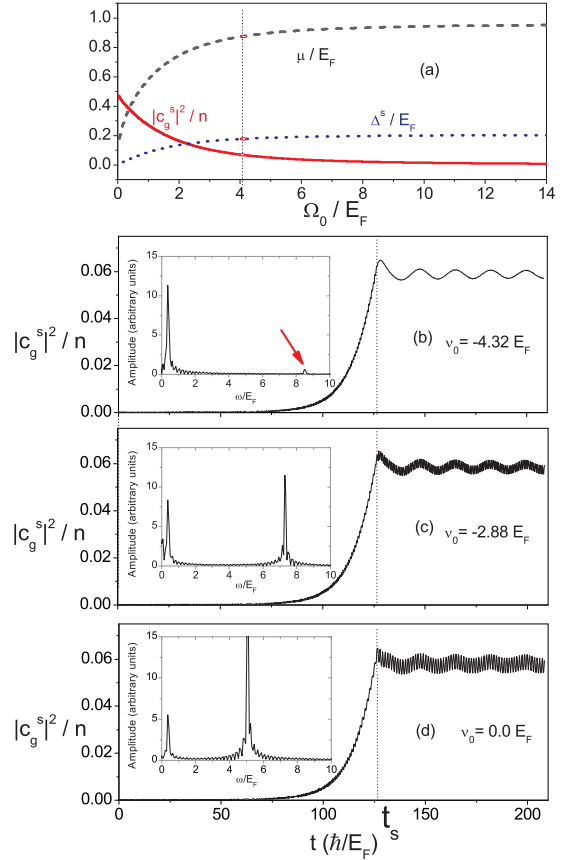}%
\caption{(Color online) (a) $\left\vert c_{g}^{s}\right\vert ^{2}/n$, $\mu$,
and $\Delta^{s}$ as functions of $\Omega_{0}$. The dark state solution at
$\Omega_{0}=4.1\ E_{F}$ is indicated by the vertical line where $\left\vert
c_{g}^{s}\right\vert ^{2}/n=0.067$, $\mu=0.87\ E_{F}$, and $\Delta
^{s}=0.17\ E_{F}$. (b)-(d) The ground population dynamics where (b) $\nu
_{0}=-4.32$ $E_{F}$, (c) $\nu_{0}=-2.88\ E_{F}$, and (d) $\nu_{0}=0.00\ E_{F}%
$. \ The insets are the Fourier spectra of the corresponding population
dynamics after $t=t_{s}=126.65\hbar/E_{F}.$ We have used the following
parameters: $U_{0}=-28.39\,E_{F}\,/k_{F}^{3}$, $g_{0}=-15.68\,E_{F}%
\,/k_{F}^{3/2}$, $n=0.034\,k_{F}^{3}$, and $K_{c}=14.4k_{F}$, where $E_{F}$
and $k_{F}$ are the fermi energy and momentum, respectively. }%
\label{Fig:populationsDynamics}%
\end{center}
\end{figure}
%EndExpansion

The stationary solutions to Eqs.~(\ref{mean-field equations}) have the form:
\begin{align*}
\label{Fig:populationDynamics}c_{m,g}\left(  t\right)   &  =c_{m,g}%
^{s}\,e^{-2i\mu t/\hbar}\,,\;\;\Delta\left(  t\right)  =\Delta^{s}\,e^{-i2\mu
t/\hbar}\,,\\
u_{k}\left(  t\right)   &  =u_{k}^{s}\,e^{iE_{k}t/\hbar}\,e^{i\mu t/\hbar
}\,,\;\;v_{k}\left(  t\right)  =v_{k}^{s}\,e^{iE_{k}t/\hbar}\,e^{-i\mu
t/\hbar}\,,
\end{align*}
where quantities with superscript $s$ are time-independent. Inserting this
stationary ansatz into Eqs.~(\ref{mean-field equations}) and searching for
solutions with $c_{m}^{s}=0$, we find that such a dark-state solution indeed
exists as long as the generalized two-photon resonance condition
\end{subequations}
\begin{equation}
\delta_{0}+\nu_{0}=2\mu\,, \label{two-photon condition}%
\end{equation}
is satisfied. Such a solution is given by $\left\vert u_{k}^{s}\right\vert
^{2}=1-\left\vert v_{k}^{s}\right\vert ^{2}=\left(  E_{k}+\epsilon_{k}%
-\mu\right)  /2E_{k},$ $E_{k}=\sqrt{\left(  \epsilon_{k}-\mu\right)
^{2}+\left\vert \Delta^{s}\right\vert ^{2}\varphi_{k}^{2}}$, where $\mu
,\Delta^{s}$ and $c_{g}^{s}$ are determined from the following equations,
representing, respectively, (a) the destructive interference condition leading
to vanishing population in $|m\rangle$
\begin{equation}
\frac{\Omega_{0}}{2}c_{g}^{s}=\frac{g}{U}\Delta^{s}\,, \label{cpt}%
\end{equation}
(b) the gap equation
\begin{equation}
\frac{1}{U}=-\frac{1}{2\pi^{2}}\int_{0}^{\infty}\frac{\varphi_{k}^{2}}{2E_{k}%
}k^{2}dk\,, \label{gap}%
\end{equation}
and (c) the conservation of particle number
\begin{equation}
n=\frac{N}{V}=2\left\vert c_{g}^{s}\right\vert ^{2}+\frac{1}{2\pi^{2}}\int
_{0}^{\infty}\left(  1-\frac{\epsilon_{k}-\mu}{E_{k}}\right)  k^{2}dk\,.
\label{number conservation}%
\end{equation}
Equation (\ref{cpt}) in particular demonstrates the coherent nature of the
dark state: for a normal atomic Fermi gas ($\Delta^{s}=0$) which does not
possess phase coherence, such a state is impossible as Eq.~(\ref{cpt}) would
imply vanishing population in the molecular level $|g\rangle$ ($c_{g}^{s}=0$).

An example of the dark state solution obtained by solving Eqs.~(\ref{cpt}%
-\ref{number conservation}) self-consistently is shown in
Fig.~\ref{Fig:populationsDynamics}(a). To remove the ultraviolet divergence in
the gap equation (\ref{gap}), we have followed the standard renormalization
procedure to replace $U$ by $\Gamma U_{0}$, where $U_{0}$ is the physical
two-body atomic collisional strength. Here $\Gamma=1/(1+U_{0}U_{c}^{-1})$, and
$U_{c}^{-1}=-mK_{c}/\left(  4\pi^{3/2}\hbar^{2}\right)  $
\cite{renormalization}. Further, by replacing $g$ with $\Gamma g_{0}$ while
keeping the rest of parameters unchanged, we can easily show that our results
become independent of $K_{c}$. Figure ~\ref{Fig:populationsDynamics}(a)
displays the ground molecular population of the dark state $|c_{g}^{s}|^{2}$,
the corresponding chemical potential $\mu$, and the gap parameter $\Delta^{s}$
as a function of the bound-bound coupling strength $\Omega_{0}$. In the limit
$\Omega_{0}/(g_{0}\sqrt{n})\rightarrow\infty$, we have $\left\vert c_{g}%
^{s}\right\vert ^{2}\rightarrow0$ and all the population is in a pure BCS
atomic state; while in the opposite limit of $\Omega_{0}/(g_{0}\sqrt
{n})\rightarrow0$, $\left\vert c_{g}^{s}\right\vert ^{2}\rightarrow0.5$ and
all the population are in the ground molecular state. Thus, in principle, we
can adiabatically convert the BCS atom pairs into ground molecular BEC or vice
versa by controlling the ratio $\Omega_{0}/g_{0}\sqrt{n}$ in the spirit of
STIRAP \cite{bergmann98}.

Our use of STIRAP here is, however, for preparing a superposition which is a
prerequisite for demonstrating coherent oscillations in fermionic systems
\cite{yuzbashyan06,andreev04,barankov04,burnett05}. Starting from $t=0$ with a
pure atomic BCS state at a relatively large $\Omega_{0}$, we adiabatically
decrease $\Omega_{0}$ to 4.1 $E_{F}$ at $t=t_{s}$ [indicated in Fig.
\ref{Fig:populationsDynamics}(b)-(d)] while maintaining the two-photon
resonance condition (\ref{two-photon condition}) through a proper chirping of
the laser frequency \cite{hong}. At $t=t_{s}$, a dark state, which is
indicated by the vertical lines in Fig.~\ref{Fig:populationsDynamics}, is then
formed with about 14\% of the atoms now converted to ground molecules. Next,
immediately after $t=t_{s}$, we suddenly change $\Omega_{0}$ from 4.1 to 4.6
$E_{F}$ and then keep it fixed for later time, while fixing all other
parameters at their respective values at $t_{s}$. The dynamical response of
the system is illustrated in Fig.~\ref{Fig:populationsDynamics}(b)-(d), which
display the ground molecular population as a function of time as obtained by
solving Eqs.~(\ref{mean-field equations}).

From the dynamical simulation, we see that the system follows the dark-state
solution up to $t=t_{s}$, after which, the sudden change of $\Omega_{0}$
induces oscillations in the population. Note that although the dark-state
solution is not explicitly dependent upon the detunings $\delta_{0}$ and
$\nu_{0}$, which must satisfy Eq. (\ref{two-photon condition}), the population
dynamics for $t>t_{s}$ does depend on their specific values. Several
conclusions can be drawn from Fig.~\ref{Fig:populationsDynamics}(b)-(d).
First, the atom-molecule dark state is robust as, after a sudden
\textquotedblleft shake\textquotedblright\ at $t_{s}$, the system oscillates
around its steady state. Second, the population oscillation occurs between the
ground molecular state $|g\rangle$ and the atomic state, while the excited
molecular population (not shown in the figures) remains negligible. Third, the
oscillations are dominated by two frequencies whose values depend on the
detunings as indicated by the corresponding Fourier spectra shown in the insets.

To better understand these oscillations and gain insight into the dark states,
we calculate the collective mode frequencies by linearizing
Eqs.~(\ref{mean-field equations}) around the dark state solution. This
procedure leads to a transcendental equation for the collective mode frequency
$\omega$%
%TCIMACRO{\TeXButton{begin widetext}{\begin{widetext}}}%
%BeginExpansion
\begin{widetext}%
%EndExpansion%
\begin{equation}
f\left(  \omega\right)  \equiv\det\left\vert
\begin{array}
[c]{cc}%
\frac{1}{U_{eff}\left(  \omega\right)  }-\int_{0}^{\infty}\frac{dk}{2\pi^{2}%
}k^{2}\varphi_{k}^{2}\frac{E_{k}^{2}+\left(  \epsilon_{k}-\mu\right)
^{2}+\omega\left(  \epsilon_{k}-\mu\right)  }{E_{k}\left(  \omega^{2}%
-4E_{k}^{2}\right)  } & \int_{0}^{\infty}\frac{dk}{2\pi^{2}}k^{2}\varphi
_{k}^{2}\frac{\left(  \varphi_{k}\Delta^{s}\right)  ^{2}}{E_{k}\left(
\omega^{2}-4E_{k}^{2}\right)  }\\
\int_{0}^{\infty}\frac{dk}{2\pi^{2}}k^{2}\varphi_{k}^{2}\frac{\left(
\varphi_{k}\Delta^{s}\right)  ^{2}}{E_{k}\left(  \omega^{2}-4E_{k}^{2}\right)
} & \frac{1}{U_{eff}\left(  -\omega\right)  }-\int_{0}^{\infty}\frac{dk}%
{2\pi^{2}}k^{2}\varphi_{k}^{2}\frac{E_{k}^{2}+\left(  \epsilon_{k}-\mu\right)
^{2}-\omega\left(  \epsilon_{k}-\mu\right)  }{E_{k}\left(  \omega^{2}%
-4E_{k}^{2}\right)  }%
\end{array}
\right\vert =0\,, \label{f(omega)=0}%
\end{equation}%
%TCIMACRO{\TeXButton{end widetext}{\end{widetext}} }%
%BeginExpansion
\end{widetext}
%EndExpansion
where $U_{eff}\left(  \omega\right)  =U+\omega g^{2}\left[  \omega\left(
\omega+2\mu-\nu_{0}\right)  -|\Omega_{0}|^{2}/4\right]  ^{-1}$. \ Here, the
integrals in the diagonal elements are automatically renormalized since
$U_{eff}\left(  \omega\right)  $ scales as $\Gamma U_{eff}^{0}$ $\left(
\omega\right)  $, where $U_{eff}^{0} ( \omega) =U_{0}+\omega g_{0}^{2}%
/[\omega( \omega+2\mu-\nu_{0}^{\prime} ) -\left\vert \Omega_{0}\right\vert
^{2} /4]$
%\begin{equation}
%U_{eff}^{0}\left(  \omega\right)  =U_{0}+\frac{\omega g_{0}^{2}}{\omega\left(
%\omega+2\mu-\nu_{0}^{\prime}\right)  -\left\vert \Omega_{0}\right\vert ^{2}%
%/4}\,,\label{U0}%
%\end{equation}
with $\nu_{0}^{\prime}=\nu_{0}+\Gamma g_{0}^{2}/U_{c}$.

Before examining $f\left(  \omega\right)  $ in detail, we first make a remark.
As we have mentioned, our dark state reduces to a pure BCS state in the limit
$\Omega_{0}/g_{0}\sqrt{n}\rightarrow\infty$. In this case, $U_{eff}\rightarrow
U_{0}$, which is independent of $\omega$. As is known \cite{volkov74}, the
collective excitation spectrum of a BCS state contains a continuous part and a
discrete mode lying just below the continuum threshold at $2\Delta^{s}$. \ Due
to the coupling between discrete (molecular) states and the continuum (atomic)
states, the problem at hand bears much resemblance to the energy
diagonalization of the Fano-Anderson type of Hamiltonians in linear atomic and
condensed matter systems \cite{fano61}. In analogy to these problems, such
discrete-continuum coupling may lead to drastic modifications to both parts of
the excitation spectrum. Mathematically, this coupling gives rise to $\omega
$-dependence in $U_{eff}$ and introduces extra poles in $f(\omega)$.

\begin{figure}
[ptb]
\begin{center}
\includegraphics[
height=3.2283in,
width=3.7386in
]%
{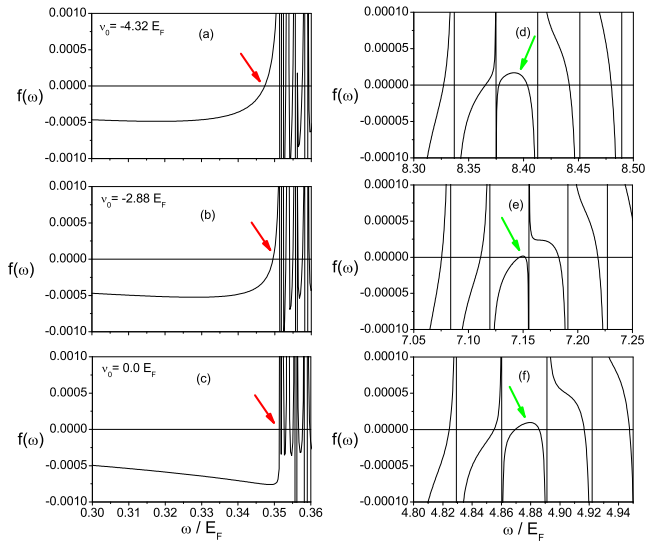}%
\caption{(Color online) The sections of $f\left(  \omega\right)  $ containing
the low frequency root (left column) and the high frequency root (right
column). \ (a) and (d) are for $\nu_{0}=-4.32\ E_{F}$, (b) and (e) are for
$\nu_{0}=-2.88\ E_{F}$, and (c) and (f) are for $\nu_{0}=0.00~E_{F}$. \ Other
parameters are the same as in Fig. \ref{Fig:populationsDynamics}.}%
\label{Fig:collectiveExcitations}%
\end{center}
\end{figure}

We now examine the spectrum by finding the roots of Eq.~(\ref{f(omega)=0}).
Since $f(\omega)$ is an even function of $\omega$, we only concentrate on the
positive-frequency branch. The function of $f(\omega)$ is plotted in
Fig.~\ref{Fig:collectiveExcitations}. The left panel
[Fig.~\ref{Fig:collectiveExcitations}(a)-(c)] shows the low-frequency part.
Here, just as in the pure BCS model, one isolated mode lies not far below the
continuum threshold. As the free-bound detuning becomes more negative, this
mode decreases and shifts further away from continuum. In the right panel
[Fig.~\ref{Fig:collectiveExcitations}(d)-(f)], we show the high-frequency
part. Here, the vertical lines are the poles determined by $\omega=2E_{k}$ at
discrete momenta. Typically, a single root is trapped between two adjacent
poles. These roots will form a continuum. This pattern of root distribution
is, however, broken in the region indicated by the arrow, where two roots
exist between two adjacent poles. In the continuous $k$ limit, one of the two
roots joins the continuum while the other one becomes part of the discrete
spectrum. The two discrete modes (one shown in left and the other in right
panel) are the ones that determine the dynamical population oscillation shown
in Fig.~\ref{Fig:populationsDynamics}(b)-(d), while the contribution from the
continous part of the spectrum, due to the destructive interference, may lead
to power-law decay of the oscillation at a longer time scale
\cite{volkov74,yuzbashyan06}.%
%TCIMACRO{\FRAME{ftbpFU}{3.7386in}{3.2283in}{0pt}{\Qcb{(Color online) The
%sections of $f\left(  \omega\right)  $ containing the low frequency root (left
%column) and the high frequency root (right column). \ (a) and (d) are for
%$\nu_{0}=-4.32\ E_{F}$, (b) and (e) are for $\nu_{0}=-2.88\ E_{F}$, and (c)
%and (f) are for $\nu_{0}=0.00~E_{F}$. \ Other parameters are the same as in
%Fig. \ref{Fig:populationsDynamics}.}}{\Qlb{Fig:collectiveExcitations}%
%}{collectiveexcitationsnew.eps}{\special{ language "Scientific Word";
%type "GRAPHIC";  maintain-aspect-ratio TRUE;  display "USEDEF";
%valid_file "F";  width 3.7386in;  height 3.2283in;  depth 0pt;
%original-width 4.1779in;  original-height 3.6045in;  cropleft "0";
%filename 'collectiveExcitationsNew.eps';file-properties "XNPEU";}}}%
%BeginExpansion
%EndExpansion

In summary, we have shown that it is possible to construct a macroscopic
atom-molecule dark state in a fermionic superfluid. The superfluidity of the
fermionic atoms is a necessary ingredient for such a state. Therefore
characteristics of the dark state may serve as a diagnostic tool for Fermi
superfluids. Via direct dynamical simulation, we have shown that the dark
state is quite robust. By perturbing the state, we are able to generate
coherent oscillations reminiscent of the oscillating current across Josephson
junctions. A remarkable feature here is that the population oscillation occurs
between the ground molecules and the BCS atom pairs, while the excited
molecular population remains highly suppressed. This has the obvious advantage
of preserving the atom-molecule coherence for a time much longer than the
excited molecular lifetime. Thus, this technique has the potential to increase
the sensibility in interference-based high-precision measurements. In
particular, the low frequency mode is directly related to the gap parameter
$\Delta^{s}$ and measurement of this frequency will allow us to gain insight
into the atom-atom and atom-molecule interactions as they will strongly affect
$\Delta^{s}$.

This work is supported by the NSF (HYL, HP), ARO (HYL), the Welch and the Keck
Foundations (HP), and by the National Natural Science Foundation of China
under Grant No. 10474055 and No. 10588402, the National Basic Research Program
of China (973 Program) under Grant No. 2006CB921104, the Science and
Technology Commission of Shanghai Municipality under Grant No. 05PJ14038, No.
06JC14026 and No. 04DZ14009 (WZ).

$\dag$To whom correspondence should be addressed E-mail: ling@rowan.edu%

\bigskip
\end{document}